# Perforation Effect on a Rectangular Metal Hydride Tank for the Hydriding and Dehydriding Process by Using COMSOL Multiphysics Software.

E.I Gkanas[1,2*], S.S Makridis[1,2], E.S Kikkinides[1] A.K Stubos[2]

1. Materials for Energy Applications Group, Department of Mechanical Engineering, University of Western Macedonia, Bacola and Sialvera Street, Kozani, 50100, Greece

2. Environmental Technology Laboratory, Institute of Nuclear Technology and Radiation Protection, NCSR "Demokritos", Agia Paraskevi, Athens, 15310, Greece

[*] Corresponding author: Bacola and Sialvera Street, Kozani, 50100, Greece, egkanas@uowm.gr

**Abstract:** In this paper, a 3-Dimensional dynamic model of a perforated rectangular metal hydride tank is presented. The metal hydride tank consists of powder $LaNi_5$ and the tubes are organized in the geometry of the rectangular in order to simulate the flux of the ambient air through the reactor, which affect hardly both the hydriding and the dehydriding reaction. A simulating study is made by solving simultaneously the energy, mass and momentum differential equations of conservation by using Comsol Multiphysics (version 4.2) software. The simulation results show great agreement with the experimental data.

1. **Introduction**

Nowadays, fossil fuel has to be replaced due to the lack of reserves. There is a number of possible energy sources, such as solar, wind, ocean currents, hydropower geothermal and others [1]. Between all these candidates, hydrogen seems to be the most promising material for energy storage because it's great properties such as regarding convenience for transportation versatility, safety and environmental compatibility [2]. Hydrogen occupies an enormous volume in normal conditions (1kg of hydrogen at ambient temperature and atmospheric pressure takes a volume of 1 $m^3$) [3]. Hydrogen also has the highest energy per unit mass and it's the most abundant element in the universe. However, due to low energy density per unit volume under normal atmospheric pressures, effective storage in limited spaces is one of the most significant technical barriers, hindering the use of hydrogen in automotive and other mobile applications [4].

$AB_5$ intermetallic compounds and especially $LaNi_5$ metal hydride are considered very promising for reversible hydrogen storage applications, due to its high storage volumetric density and operating conditions of pressure and temperature compatible with fuel cells (T = 20-100 $^0C$ and p = 1-15 bar) [5]. Also, these alloys have fairly flat plateaus and small hysteresis [6] and consider to have inferior storage properties [7, 8].

The basic phenomena which underlie hydrogen absorption and desorption in metal hydrides have long been studied with special focus on their kinetic characteristics which have analyzed by means of experimental investigation and numerical models [9-17].

In the current work, a simulation study of both the hydriding and dehydriding process is made, in a 3 – Dimensional metal hydride tank which contains LaNi5 alloy. A comparison is made between a perforated rectangular metal hydride tank and a conventional rectangular and the effect of the presence of ambient air inside the tubes is studied in order to improve the storage of hydrogen.

Mathematical simulation have been achieved through the description of transport phenomena by partial differential equations by using Comsol Multiphysics (version 4.2). We simultaneously solved the differential equations of Energy, Mass and Momentum conservation laws, taking into account the proper initial and boundary conditions.

## 2. Numerical model

The geometry of the metal tank is seen in Figure 1. The hydride bed is filled with powdered LaNi$_5$, and the tank is perforated, so that ambient air in constant temperature is passing through the tubes and cooling the tank. The main geometrical and thermal – physical parameters used in the simulation are reported on Table 1.

In order to simplify the model, some assumptions has been made. The main assumptions considering for developing the model are the following:

The media (metal and hydrogen) are in local thermal equilibrium (the gas temperature is the same as of the solid temperature)

The solid phase is isotropic and has a uniform porosity

Hydrogen is treated as an ideal gas, from a thermodynamic point of view.

The effect of hydrogen concentration on the variation of equilibrium pressure is negligible (Van't Hoff law)

Thermal – physical properties are constant

The heat transfer by radiation is neglected. This assumption is valid for all Mm, La, Zr and Ti based alloys, whose minimum absorption temperatures are well below 30$^0$C.

The governing equations, consist of energy, mass and momentum conservation which described by partial differential equations, and some other equations that describe the kinetics of absorption and desorption.

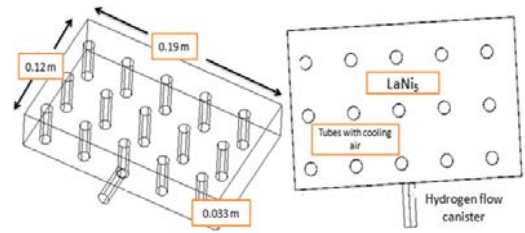

**Figure1** Geometry of the simulation domains. The model consist of three main subdomains, the perforated rectangular metal hydride tank (subdomain 1), the tubes (subdomain 2), and the hydrogen supply canister (sundomain 3). The dimensions of the rectangular are: 0.19x0.12x0.033

### 2.1 Energy equation

Assuming thermal equilibrium between the hydride powder and hydrogen, a single energy equation is solved instead of separate equations foe both solid and gaseous phases.

$$(\rho \cdot Cp)_e \cdot \frac{\partial T}{\partial t} + (\rho_g \cdot Cp_g) \cdot \bar{v}_g \cdot \nabla T$$
$$= \nabla \cdot (k_e \cdot \nabla T) + m \cdot (\Delta H - T \cdot (Cp_g - Cp_s))$$
(1)

Considering only parallel heat conduction in solid and gas phases, there are the following expressions for specific heat and thermal conductivity respectively:

$$(\rho \cdot Cp)_e = (\varepsilon \cdot \rho_g \cdot Cp_g) + ((1-\varepsilon) \cdot \rho_s \cdot Cp_s) \quad (2)$$

$$k_e = \varepsilon \cdot k_g + (1-\varepsilon) \cdot k_s \quad (3)$$

Both the equations (2) and (3) are expressed as porosity – weighted functions of the hydrogen – gas and the solid – metal phases.

### 2.2 Hydride mass balance

For the solid, a mass conservation equation is considered. $(1-\varepsilon) \cdot \frac{\partial (\rho_s)}{\partial t} = -m$ **(4)**

The mass conservation for the gas is considered as:

$$\varepsilon \cdot \frac{\partial(\rho_g)}{\partial t} + div(\rho_g \cdot \vec{v}_g) = -m \quad (5)$$

### 2.3 Momentum equation

The gas velocity can be expressed using Darcy's law. By neglecting the gravitational effect, the equation is the above:

$$\vec{v}_g = -\frac{K}{\mu_g} \cdot grad(\vec{P}_g) \quad (6)$$

Where K is the permeability of the solid and $\mu_g$ is the dynamic viscosity of gas. The solid permeability is given by the Kozeny – Carman's equation:

$$K = \frac{dp^2 \cdot \varepsilon^3}{180 \cdot (1-\varepsilon^2)} \quad (7)$$

Assuming that the hydrogen is an ideal gas, from the perfect gas law ($\rho_g = (P_g M_g)/(RT)$) and considering Darcy's law, the mass conservation equation of hydrogen becomes:

$$\frac{\varepsilon \cdot M_g}{R \cdot T} \cdot \frac{\partial P_g}{\partial t} + \frac{\varepsilon \cdot M_g \cdot P_g}{R \cdot T} \cdot \frac{\partial}{\partial t} \cdot \frac{1}{T} -$$
$$\frac{K}{v_g \cdot r} \cdot \frac{\partial}{\partial r} \cdot \frac{r \cdot \partial P_g}{\partial r} - \frac{K}{v_g} \cdot \frac{\partial^2 P_g}{\partial z^2} = -m$$
$$(8)$$

### 2.4 Kinetic expression

For the absorption and desorption of hydrogen, the following kinetic expressions are used:

$$m = C_a \cdot \exp[-\frac{E_a}{R_g \cdot T}] \cdot \ln[\frac{p_g}{P_{eq}}] \cdot (\rho_{ss} - \rho_s)$$
(9)

$$m = Cd \cdot \exp[-\frac{E_d}{R_g \cdot T}] \cdot (\frac{P_{eq} - p_g}{P_{eq}}) \cdot (\rho_s - \rho_o)$$
(10)

Where (9) is for absorption and (10) for desorption respectively, and m is the source term and used in equations (1), (4), (5), (8). $C_a$ and $C_d$ are pro-exponential constants for absorption and desorption respectively, $E_a$ and $E_d$ are the absorption/desorption activation energy, $\rho_{ss}$ is the saturation density for hydride, and $\rho_o$ is the initial metal hydride density.

### 2.5 Equilibrium pressure

The equilibrium pressure for the hydrogen, which is the most important parameter which defines if the reaction is going to take place or not, is given by van't Hoff law:

$$\ln P_{eq} = \frac{\Delta H}{R_g \cdot T} - \frac{\Delta S}{R_g} \quad (11)$$

### Initial and boundary conditions

Initially, gas and solid are at the same temperature. Pressure and hydride density are assumed to be constant.

$$T_s = T_g = T_o$$

$$p_g = p_o$$

$$\rho_s = \rho_0$$

Our geometry consists of a rectangular tank, with tubes well distributed perpendicular to the rectangular surface. In Figure 2a is seen the walls of the geometry in where there is a heat flux procedure, and the heat flux condition is set:

$$-\vec{n} \cdot k_e \cdot \nabla T = h \cdot (T - T_f) \quad (12)$$

The boundaries of the tubes are set with the condition T = T$_0$, where T$_0$ is the temperature of the cooling air. This is shown in Figure 2b.

The common boundaries between the tube and the perpendicular surface of the reactor has the following condition which corresponds to convective flux. This depicts in Figure 2c.

$$-\vec{n} \cdot k_e \cdot \nabla T = 0 \quad (13)$$

Finally, on the hydrogen supply canister (Figure 1), the boundary conditions are: Thermal insulation at the collateral walls, continuity ($-\vec{n} \cdot (k_{eq1} \cdot \nabla T_1 - k_{eq2} \cdot \nabla T_2) = 0$) has set at the common contact area of the supply canister with the tank.

| Parameter | Symbol | Value |
|---|---|---|
| **Initial hydride density** | $\rho_0$ | 8300 Kg/m$^3$ |
| **Saturation hydride density**[12,13] | $\rho_{ss}$ | 8354 Kg/m$^3$ |
| **Hydride molecular weight**[12,13] | $M_s$ | 432.45 gr/mol |
| **Hydride specific heat** | $C_{ps}$ | 355 J/kgK |
| **Hydride porosity** | $\varepsilon$ | 0.5 |
| **Reaction enthalpy**[13] | $\Delta H$ | 30000J/mol |
| **Reaction entropy**[13] | $\Delta S$ | 104.7 J/molK |
| **Heat transfer coefficient**[13] | h | 400 W/m$^2$K |
| **Hydride conductivity** | $\lambda$ | 1.32 W/mK |

Table 1. Parameters used in the model.

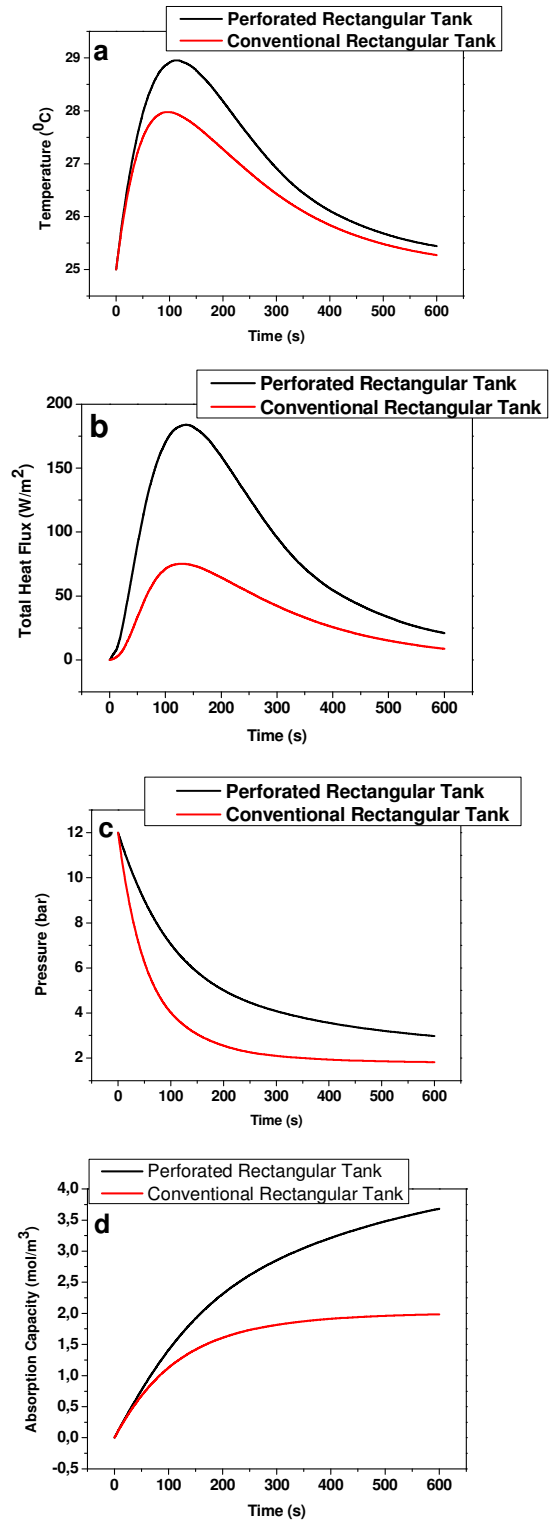

**Figure 2**. Comparison of the absorbing response for the perforated rectangular tank and the conventional rectangular tank. Figure 2a shows the temperature evolution for both beds. Figure 2b shows heat flux behavior inside the tubes. Figure 2c depicts pressure evolution with time. According van't Hoff law Peq=1.8bar. The results are the same. Figure 2d shows absorption capacity for the two beds. Perforated tank has better absorption efficiency.

## 3. Results

### 3.1 Comparison of the hydriding response between the perforated rectangular and the conventional rectangular

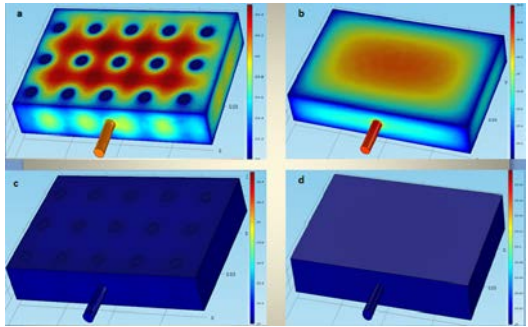

**Figure 3**. Temperature distribution at the surface of the tanks. 4a shows the temperature distribution for the perforated rectangular after 120s, and 4b for the conventional rectangular after 105s. 4c and d shows the distribution at the end of reaction (600s) for the two rectangular respectively. The external bath's temperature has achieved.

Figure 2 shows the variation of temperature, total heat flux, pressure and absorption capacity with time for the absorption process for the two metal hydride beds. The initial temperature for both beds is $25^0$C and the temperature of the cooling bath at the walls has also set at $25^0$C.

From figure 2, we assume that the presence of the tubes where ambient air passes through affects the hydriding reaction and from 2d we also observe that the storage capacity is bigger for the perforated rectangular tank. For the perforated tank the capacity is almost 4mol/m$^3$ while the capacity for the conventional tank is at 1.5 mol/m$^3$. Figure 3 also presents the temperature distribution across the surface of the tanks at their maximum temperature (after 120 and 105s respectively) and after 600s where the hydriding reaction has already end. After that time the temperature decreases until the equilibrium temperature corresponds to the bath's temperature ($25^0$C).

### 3.2 Comparison of the dehydriding response between the perforated rectangular and the conventional rectangular.

Hydrogen desorption is a high endothermic process, which mean that the hydrogen release process is not able to start automatically. An external heat source with high temperature is necessary in order the dehydriding reaction to start. In this study, the external heat source is a hot bath at $80^0$C. Figure 4 shows the evolution of temperature, total heat flux, pressure and absorption capacity with time for the desorption process for the two metal hydride beds. The effect of the presence of a cooling medium inside the tank at the desorption process has studied. The results show that the reduction of temperature via cooling inside the tank affects the reaction due to the endothermic nature of the dehydriding process. Figure 5a also presents shows the surface temperature distribution for the perforated tank after the first 15s of the reaction and 5c the temperature distribution at the end of the reaction, while 5b shows the temperature distribution for the conventional tank after the first 20s of the reaction and 5d at the end of the dehydriding process.

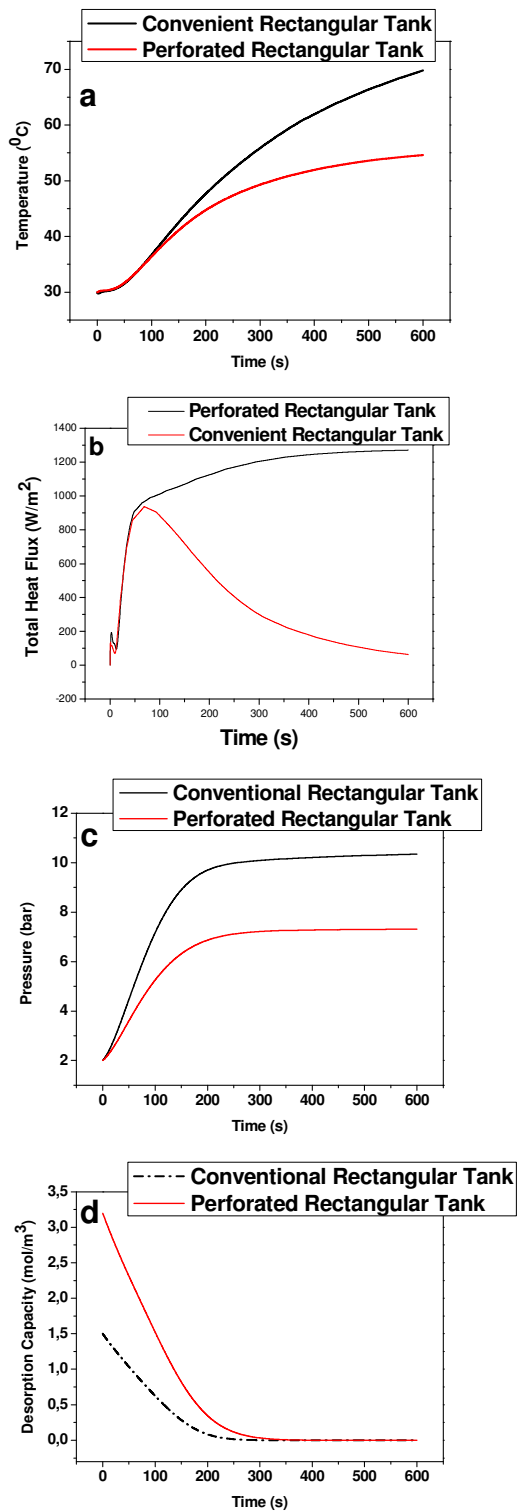

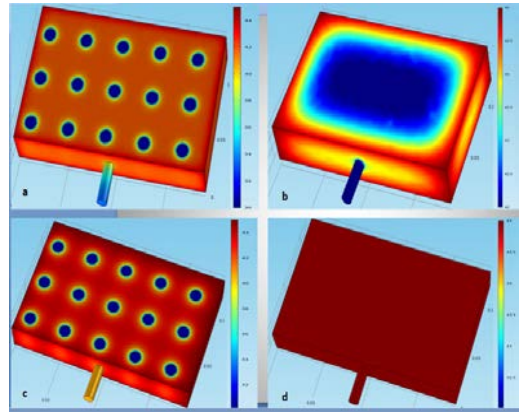

Fig 5. Temperature distribution at the surface of the tanks. 5a shows the temperature distribution for the perforated tank after 15s, and 5b for the conventional tank after 20s. 5c and d points the distribution of temperature at the end of the reaction.

**Figure 4**. Comparison of the dehydriding response for the perforated rectangular tank and the conventional rectangular tank. Figure 4b shows the total heat flux behavior inside the tanks. Figure 4c depicts pressure evolution inside tanks. According to van't Hoff law Peq = 10.5bar for BED – 2 which totally match with the results. Figure 4d shows desorption capacity for the two tanks.

4. **Conclusions.**

In the current study, simulation results for the absorption and desorption process inside two different hydride tanks and the comparison between them was presented. Results show that the presence of the tubes in the perforated metal hydride tank affects hardly both the absorption and desorption process due to the presence of a cooling medium which was ambient air. The presence of the cooling medium seems to affect positive the absorption process and for the desorption process we are able to control the reaction by reducing the reaction temperature.

5. **References.**

**6. Nomenclature**

$C_p$: Specific Heat (J/kgK)
$u$ : Gas Velocity (m/s)
$E_a$: Absorption Activation Energy (J/mol)
$E_d$: Desorption Activation Energy (J/mol)
$K$: Permeability (m$^2$)
$P$: Pressure (bar)
$T$: Temperature (K)
$t$: Time(s)
$R$: Gas Global Constant, (8.314 J/molK)
$d_p$: Particle Size (m)
Greek Symbols
$\Delta H$: Reaction Enthalpy (J/mol)
$\varepsilon$: Porosity
$\Delta S$: Reaction entropy (J/molK)
$\lambda$: Thermal Conductivity(W/mK)
$\mu_g$: Dynamic Viscosity (kg/ms)
$\rho$: Density(kg/m$^3$)
Subscripts
eq: Equilibrium
g: Gas
m: Metal